# Optical modulation in a resonant tunneling relaxation oscillator

J. M. L. Figueiredo[a)], C. R. Stanley, A. R. Boyd and C. N. Ironside

Department of Electronics and Electrical Engineering, University of Glasgow, Glasgow G12 8LT, United Kingdom

S. G. McMeekin

Cardiff School of Engineering, University of Wales Cardiff, PO Box 917, Newport Rd., Cardiff, NP2 1XH, United Kingdom

Electronic mail: jlongras@elec.gla.ac.uk

[a)]Also with the Centro de F'sica do Porto - ADFCUP, Universidade do Porto, Rua do Campo Alegre 687, 4150 PORTO, Portugal

A. M. P. Leite

Centro de F'sica do Porto - ADFCUP, Universidade do Porto, Rua do Campo Alegre 687, 4150 PORTO, Portugal

We report high speed optical modulation in a resonant tunneling relaxation oscillator consisting of a resonant tunneling diode (RTD) integrated with a unipolar optical waveguide and incorporated in a package with a coplanar waveguide transmission line. When appropriately biased, the RTD can provide wide-bandwidth electrical gain. For wavelengths near the material band-edge, small changes of the applied voltage give rise to large, high-speed electro-absorption modulation of the light. We have observed optical modulation at frequencies up to 14 GHz, associated with sub harmonic injection locking of the RTD oscillation at the fundamental mode of the coplanar transmission line, as well as generation of 33 ps optical pulses due to relaxation oscillation.

Resonant tunneling diodes (RTDs) have been widely studied because of their potential application in high frequency signal generation,[1] high speed signal processing at microwave frequencies,[2] and in optoelectronics.[3-6] Recent studies on resonant tunneling relaxation oscillators (RTROs)[7,8] have shown generation of 30 ps electrical pulses at a repetition rate of 1.1 GHz, and harmonic and sub-harmonic locking. In an optical waveguide containing a double-barrier RTD implemented in the AlGaAs material system, our group previously reported electro-absorption modulation at 900 MHz with a modulation depth of 7 dB.[9] The modulation relies on a depletion region that is formed in the waveguide core and, depending on the bias condition, a substantial part of the terminal voltage may be dropped across this region. From the I-V characteristic of the RTD optical waveguide modulator, it can be seen that small changes of bias voltage close to the peak-to-valley transition region can cause large changes in the electric field distribution across the depleted part of the waveguide core. The electric field shifts the absorption band-edge to longer wavelengths via the Franz-Keldysh effect[10] and, therefore, changing the transmission characteristics of the waveguide. Compared to the conventional pn electro-absorption modulator, the advantage of the RTD modulator is that when DC biased close to the NDR region, the device behaves as a optical waveguide electro-absorption modulator integrated with a wide bandwidth electrical amplifier. This letter reports results of high speed light modulation in a resonant tunneling relaxation oscillator configuration, consisting of a RTD optical waveguide modulator integrated with a coplanar waveguide transmission line.

The RTD optical waveguide structure was grown by Molecular Beam Epitaxy (MBE), a Varian Gen II system, on a semi-insulating GaAs substrate, figure 1.(a). It consists of two 1.4 nm thick AlAs barriers separated by a 7 nm wide GaAs quantum well, and two 500 nm thick moderately doped (Si: $2\times10^{16}$cm$^{-3}$) GaAs spacer layers each side of the double barrier, which are surrounded by heavily doped (Si: $2\times10^{18}$cm$^{-3}$) AlGaAs cladding layers, allowing light confinement in the direction parallel to the double barrier plane. A n$^+$ GaAs cap layer was provided for formation of Au-Ge-Ni ohmic contacts. The ridge waveguides (2 to 6 µm wide) and large-area mesas each side of the ridges, were fabricated by dry-etching. Ohmic contacts (100 to 400 µm long) were deposited on top of the ridges and mesas. A SiO$_2$ layer was deposited, and access contact windows were etched on the ridge and the mesa electrodes, figure 1.(b), allowing contact to be made to high frequency bonding pads. After cleaving, the devices were die bonded on packages containing 50 Ω coplanar waveguide (CPW) transmission lines of different lengths. The device pads were connected to the package CPW pads via gold wires; a SMA connector was then soldered to the package CPW line, figure 1.(c). The dc and RF signals were applied through a 40 GHz bandwidth bias tee. The dc current-voltage characteristics of the packaged devices show typical RTD behaviour, with the peak current density being approximately 13 kA cm$^{-2}$ (800 µm$^2$ active area) and a peak-to-valley current ratio around 1.5. Peak voltages were in the range 2.1 to 2.5 V and the valley voltage varied from 2.6 to 3.2 V.

The optical characterisation employed light from a Ti:sapphire laser, tuneable in the wavelength region around the absorption edge of the GaAs waveguide (850-950 nm). Light was coupled into the waveguide by a microscope objective end-fire arrangement. To measure the change in the optical absorption spectrum induced by the peak-to-valley transition, a RF signal was injected to switch the RTD between the extremes of the NDR region, and a photodetector was used to measure the transmitted light. The band edge shift was found to be approximately 12 nm. The high speed optical response of the modulator was measured with a streak camera (Hamamatsu C5680) with a minimum time resolution of around 2 ps. A part of the injected RF signal power was required to trigger the streak camera. Figure 1.(c) presents schematically the experimental set up.

Due to the highly non-linear I-V characteristic in the NDR region, the RTD is able to generate many high-order odd harmonics of the injected signal.[8] The optical modulation associated with locking of the RTD to the frequency of the fundamental mode of the CPW transmission line package was studied. Optical modulation at 14 GHz was observed in a packaged device with a 5 mm long transmission line, dc biased in the NDR region (2.25 V), when a RF signal of 0.4 V amplitude and frequency around 1 GHz was injected. A packaged device with a 7 mm long transmission line, again dc biased in the NDR region (2.4 V), showed optical modulation at 8 GHz for an injected signal of 0.6 V amplitude and frequency around 2 GHz. For this configuration the modulator bandwidth-to-drive-voltage ratio is higher than 33 GHz/V. The streak camera traces for the 14 GHz and 8 GHz optical responses are presented in figure 2.

When scanning the frequency of the injected signal with amplitude in the range 0.8 - 1.4 V, it was found that the RTD optical modulator produced optical pulses with a full width at half maximum (FWHM) of approximately 33 ps at frequencies which are associated with the device operation as a relaxation oscillator.[7] According to reference 7, the pulses are caused by the discharge of the RTD capacitance when the RTD switches from the second positive differential resistance (PDR) region to the first PDR region. Figure 3 displays the streak camera trace of an optical pulse with 33 ps FWHM. On a longer time scale other steak camera results indicate a pulse repetition rate around 166 MHz. The trace shows a maximum modulation depth of 18 dB, which suggests that the RTD switches between two astable points well into the PDR regions. If we follow the explanation of Ref. 7, when starting at point A (the upper dwell zone of the relaxation oscillator) the collector is fully depleted and the electric field is high, giving low transmission. A return pulse forces the trajectory down below the valley, turning up a few ps later, towards point B (the lower dwell zone). During switching, the RTD capacitance first discharges, strongly decreasing the electric field across the depleted region which gives rise to high transmission. The RTD capacitance starts to recharge causing the electric field to rise to the astable value associated with the first dwell zone. Transmission at point B is higher than at point A, because now the collector region is not fully depleted. Switching from B to A depletes the entire collector region, recharging the RTD capacitance and making the transmission drop again. This drop is not evident in figure 3 because of the short time scale operation of the streak camera required to resolve the 33 ps pulse.

In conclusion, integrating a RTD with an optical waveguide is a relatively easy way of combining a wide bandwidth electrical amplifier with an electro-absorption modulator opening up the possibility of a variety of modes of operation. A RTRO configuration has been demonstrated, providing optical modulation up to 14 GHz when a 0.4 V amplitude RF signal around 1 GHz is injected. Relaxation oscillator operation of the RTD produces optical pulses as short as 33 ps and with modulation depth up to 18 dB, for a range of amplitudes and frequencies of the injected signal. At this stage, development of the device concept appears to be a promising route towards a high speed, low power, optoelectronic modulator, with operation extended to 1550 nm possible by using the InAlGaAs quaternary system.

J.M.L. Figueiredo acknowledges FCT-PRAXIS XXI - Portugal for his Ph.D. grant.

**Figure Captions**

FIG. 1. (a) Schematic diagram of the wafer structure. (b) The RTD optical waveguide (RTD-OW) modulator configuration. (c) Diagram of a packaged device and schematic measurement set up.

FIG. 2. Streak camera traces of the modulator optical response associated with locking of the RTD-OW oscillation to the fundamental mode of a package with two different transmission line. (a) Trace corresponding to 14 GHz optical modulation for a packaged device with a transmission line 5 mm long. (b) Trace of the 8 GHz modulator optical response for a packaged device with a transmission line 7 mm long.

FIG. 3. Streak camera measurement of the 33 ps pulse optical response, showing modulation up to 18 dB.

FIG. 1., J.M.L. Figueiredo, Applied Physics Letters

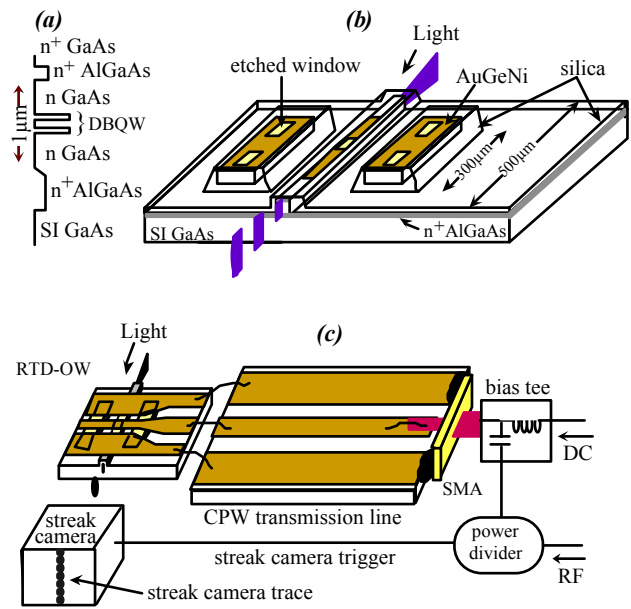

FIG. 2., J.M.L. Figueiredo, Applied Physics Letters

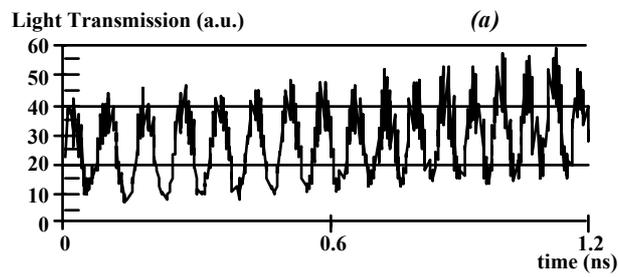

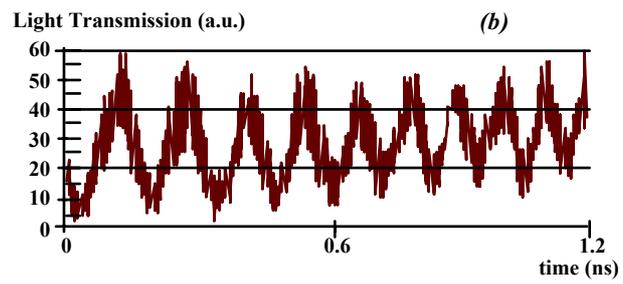

FIG. 3., J.M.L. Figueiredo, Applied Physics Letters

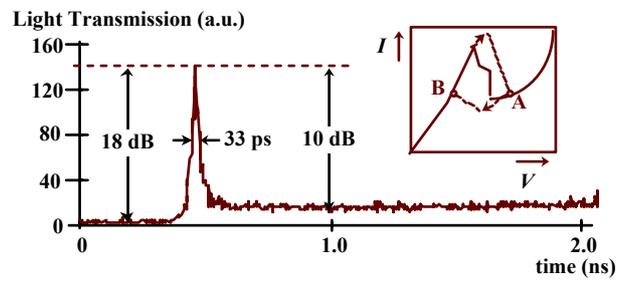